\documentclass[useAMS,usenatbib]{mn2e}
\pdfoutput=1
\usepackage[dvips, pdftex]{graphicx}
\usepackage{epstopdf}
\usepackage{amssymb}
\usepackage{amsmath}
\long\def\symbolfootnote[#1]#2{\begingroup \def\thefootnote{\fnsymbol{footnote}}\footnote[#1]{#2}\endgroup}

\title[Slowing down atomic diffusion in subdwarf B stars]{Slowing down atomic diffusion in subdwarf B stars:\\
mass loss or turbulence?}
\author[Haili Hu et al.]{Haili Hu$^{1}$\thanks{E-mail:
hailihu@ast.cam.ac.uk}, C.~A. Tout$^{1}$, E. Glebbeek$^{2}$ and M.-A.~Dupret$^{3}$\\
$^{1}$Institute of Astronomy, The Observatories, Madingley Road, Cambridge CB3 0HA\\
$^{2}$Department of Physics \& Astronomy, McMaster University, 1280 Main Street West, Hamilton, Ontario, Canada L8S 4M1\\
$^{3}$Institute d'Astrophysique et G\'eophysique de l'Universit\'e de Li\`ege, All\'ee du 6 Ao\^ut 17, 4000 Li\`ege, Belgium  }
\begin{document}

\date{Accepted 2011 July 20.  Received 2011 July 19; in original form 2011 June 14}

\pagerange{\pageref{firstpage}--\pageref{lastpage}} \pubyear{2011}

\maketitle

\begin{abstract}
%Context
Subdwarf B stars show chemical peculiarities that cannot be explained by diffusion theory alone. Both mass loss and turbulence have been invoked to slow down atomic diffusion in order to match observed abundances. The fact that some sdB stars show pulsations gives upper limits on the amount of mass loss and turbulent mixing allowed.  
%Aims
Consequently, non-adiabatic asteroseismology has the potential to decide which process is responsible for the abundance anomalies. 
%Methods
We compute for the first time seismic properties of sdB models with atomic diffusion included consistently during the stellar evolution. The diffusion equations with radiative forces are solved for H, He, C, N, O, Ne, Mg, Fe and Ni. We examine the effects of various mass-loss rates and mixed surface masses on the abundances and mode stability.  
%Results
It is shown that the mass-loss rates needed to simulate the observed He abundances ($10^{-14}\lesssim\dot{M}/{\rm M}_{\odot}\,{\rm yr}^{-1}\lesssim10^{-13}$) are not consistent with observed pulsations. We find that for pulsations to be driven the rates should be $\dot{M}\lesssim10^{-15}\,{\rm M}_{\odot}\,{\rm yr}^{-1}$. On the other hand, weak turbulent mixing of the outer $10^{-6}\,{\rm M}_{\odot}$ can explain the He abundance anomalies while still allowing pulsations to be driven. The origin of the turbulence remains unknown but the presence of pulsations gives tight constraints on the underlying turbulence model. \end{abstract}


\section{Introduction}
Subdwarf B (sdB) stars are subluminous B-type stars that show strong broad Balmer lines  and weak or no He I absorption lines. They are ubiquitous not only in our own Galaxy, where they are found in all stellar populations, but also in old galaxies where they are believed to cause the UV-upturn \citep{brown1997}. \citet{heber1986} linked the sdB evolutionary stage to  that of Extreme Horizontal Branch (EHB) stars which are core He burning stars with an unusual thin H envelope ($M_{\rm env}<0.02$ M$_{\odot}$). Such a star must have formed either from a red giant that lost nearly its entire H envelope or from a He white dwarf merger. Although the formation channels are well studied (\citealt{han2003} and references therein), they cannot explain all observed features such as binary period distribution and slow rotation. A promising way to discriminate between different formation channels is with asteroseismology \citep{hu2008} but, before this can be achieved, the theoretical models need improving. Currently, the uncertainties in the   models are larger than the observational errors obtained with space missions such as CoRoT and \emph{Kepler}. 
This paper is part of our ongoing work \citep{hu2009,hu2010} to improve the input physics used in seismic modelling of sdB stars.

Interestingly, two subgroups of sdB stars are pulsating with variable class names V361\,Hya and V1093\,Her. The discovery of short-period ($100-400$ s) pulsations in V361\,Hya stars by \citet{kilkenny1997} coincided with the prediction of unstable pressure($p$)-mode pulsations in sdB stars by \citet{charpinet1996}. The excitation was attributed to the opacity mechanism enabled by Fe accumulating diffusively in the stellar envelope \citep{charpinet1997}.
A few years later,  \citet{green2003} discovered long-period pulsations ($30-120$ min) in the cooler V1093\,Her stars. \citet{fontaine2003} showed that the same opacity mechanism could also excite long-period gravity($g$)-modes. However, they found too cool a theoretical blue-edge of the $g$-mode's instability strip  compared with observations. The problem can be partly solved by using OP opacities \citep{badnell2005} instead of OPAL's \citep{iglesias1996} \`and by assuming that Ni accumulates as well as Fe  \citep{jeffery2006}. Furthermore, inclusion of H-He diffusion shifts the theoretical blue-edge even closer to the observed value of $30\,{\rm kK}$ \citep{hu2009}. It is evident that correct modelling of the V1093\,Her instability strip awaits seismic models with time-dependent diffusion such as we produce here.

Also interesting is that sdB stars are chemically peculiar without significant spectroscopic differences between variable and constant stars \citep{otoole2006,blanchette2008}. Typically He is deficient with number ratios $10^{-4}\leq n_{\rm{He}}/n_{\rm{H}} \leq0.1$ and there is a trend for He abundance to increase with $T_{\rm{eff}}$ \citep{edelmann2003}.  The Fe abundance is close to solar irrespective of stellar population and $T_{\rm{eff}}$ whereas other iron-group elements, such as Ni, can be strongly enriched. Light metals (C, N, O, Ne, Mg) are usually depleted, although there is some scatter between stars \citep{edelmann2006,geier2010}. The observed abundance anomalies can be explained in part by atomic diffusion acting in the radiative envelope. Atomic diffusion alone, however, would cause all surface He to sink on a very short time-scale (about $10^4$\,yr) compared to the EHB evolutionary time-scale (about $10^8$\,yr). Hence there must be a competing process which is generally accepted to be mass loss. Recently turbulence has been proposed to play this role instead \citep{michaud2011}.

The observed He abundances have been reproduced with models that include atomic diffusion and mass-loss rates of $10^{-14}-10^{-13}\, {\rm M}_{\odot}\,{\rm yr}^{-1}$  \citep{fontaine1997, unglaub2001}. For higher rates the effects of diffusion become unnoticeable, whereas for lower rates He would sink too quickly. \citet{vink2002} presented a mass-loss recipe for (E)HB stars derived from line-driven wind models. {Their results give upper limits for the mass loss rates ($\dot{M}<10^{-11}\, {\rm M}_{\odot}\,{\rm yr}^{-1}$) and might apply to the most luminous sdB stars that show anomalous H$\alpha$ line profiles \citep{heber2003,vink2004}. However, for  most sdBs no observational evidence of mass loss has been found so far.} An independent wind model by \citet{unglaub2008} showed that for weak winds, $\dot{M}\lesssim 10^{-12}\,{\rm M}_{\odot}\,{\rm yr}^{-1}$, the metals decouple from H and He and, for rates below $10^{-16}\,{\rm M}_{\odot}\,{\rm yr}^{-1}$, the winds become purely metallic. Such selective winds could be responsible for some of the observed metal abundance anomalies but they cannot explain the observed He abundances.

Alternatively, \citet{michaud2011} showed that turbulent mixing of the outer $10^{-7}\,{\rm M}_{\odot}$ could also produce most of the observed abundance anomalies. Unfortunately, their computational setup restricted their calculations to the first $32\,{\rm Myr}$ of the E(HB) and to He abundances above $X({\rm He})\approx 10^{-4}$. They do not give a physical origin for the turbulence but simply mix the outer layer in order to reproduce the solar Fe abundance observed in most sdB stars. In their models Fe reaches solar abundance in the entire mixed region. This could prevent the driving of pulsation modes. 

Also mass loss works against the driving mechanism. According to  \citet{chayer2004} and \citet{fontaine2006} mass-loss rates above  $\dot{M}\approx6\times10^{-15}\,{\rm M}_{\odot}\,{\rm yr}^{-1}$ would gradually destroy the Fe reservoir and an sdB pulsator becomes constant within $10^7\,{\rm yr}$ as it evolves. Their results are based on a non-diffusive stellar model assuming an initial Fe profile from an equilibrium between gravitational settling and radiative levitation. The question is, however, can Fe build up in the first place in the presence of stellar winds? If so, shouldn't diffusion take place at the same time that mass is lost  and help build up the Fe reservoir again? To answer these questions we perform a consistent analysis where we compute the effects of mass loss and atomic diffusion simultaneously. Furthermore, we determine which process (mass loss or turbulence) is responsible for retarding atomic diffusion in sdB stars by {{performing a pulsational stability analysis. Since non-adiabatic effects are responsible for mode driving and damping, such a study is termed non-adiabatic asteroseismology.} 

Beware that the term `atomic diffusion'  is not always used consistently in the literature. For clarification, in this work we include all physical processes that contribute to atomic diffusion. These are gravitational settling, thermal diffusion, concentration diffusion and radiative levitation. The corresponding abundance changes are caused by pressure gradients, temperature gradients, concentration gradients and radiative forces, respectively. Indeed, we treat atomic diffusion in a multi-component fluid \citep{burgers1969}, using diffusion coefficients derived from a screened Coulomb potential \citep{paquette1986}. We account for partial ionization by using a mean ion charge per element. We calculate radiative forces from first principles using atomic data. Our  approach gives a high accuracy treatment of atomic diffusion because it is free of many common assumptions, such as full ionization, simplified treatment of trace elements and approximate radiative forces.

We explain our method in Sect.~\ref{method}. In particular the computations of stellar evolution (Sect.~\ref{evcode}), radiative accelerations (Sect.~\ref{OP}) and atomic diffusion (Sect.~\ref{burgers}) are described. In Sect.~\ref{results}, we present the results for typical sdB models. The effects of atomic diffusion (Sect.~\ref{diffusion}), mass loss (Sect.~ \ref{massloss}) and turbulence (Sect.~\ref{turbulence}) on the abundances and mode stability are evaluated. We summarize and discuss our main results in Sect.~\ref{summary}. We also give a brief discussion of the possible physical origin for turbulence.

\section{Computational method}\label{method}
This work presents the first seismic computations of stellar models with a self-consistent treatment of atomic diffusion. The changes in the chemical abundances are accounted for by continuously updating the radiative accelerations and the Rosseland mean opacity, not only during the stellar evolution but also in the pulsation calculations. We describe here the computational tools and methods we use to accomplish this.

\subsection{The stellar evolution code}\label{evcode}
We compute stellar models with the stellar evolution code STARS \citep{eggleton1971}. This code has been frequently updated (e.g.~\citealt{pols1995}) and many derivatives are circulating. In our version the necessary modifications have been made to construct stellar models suitable for seismic studies with the non-adiabatic pulsation code MAD written by \citet{dupret2001} (see \citealt{hu2008}). Furthermore, the implementation of gravitational settling, thermal diffusion and concentration diffusion is described in \citet{hu2010}. In the present work we add the process of radiative levitation. We follow the abundance changes of H, He, C, N, O, Ne, Mg, Fe and Ni while the remaining minor species are kept constant. 

\subsubsection{Input physics}
We take nuclear reaction rates from \citet{angulo1999}, except for the $^{14}{\rm N}({\rm p}, \gamma)^{15}{\rm O}$ rate for which we use the recommended value by \citet{herwig2006} and \citet{formicola2002}. Neutrino loss rates are according to \citet{itoh1989,itoh1992}. Convection is treated with a standard mixing-length prescription \citep{bohm-vitense1958} with a mixing length to pressure scale height ratio of $l/H_p=2.0$. Convective mixing is treated as a diffusive process in the framework of mixing length theory. 

The Rosseland mean opacity $\kappa_{\rm{R}}$ is computed in-line during the evolution to account for composition changes consistently. For this we use data and codes from the Opacity Project (OP, \citealt{seaton1994} and \citealt{badnell2005}) which also allows computation of radiative accelerations $g_{\rm rad}$ (see Section \ref{OP}). This is in contrast to \emph{all} previous work with the STARS code that made use of interpolation in pre-built opacity tables. For high temperatures outside the OP range ($T>10^8$ K), we still use OPAL tables   \citep{iglesias1996} combined with conductive opacities \citep{cassisi2007}.

\subsubsection{Simultaneous solution}
The STARS code distinguishes itself from other evolution codes by its fully simultaneous, implicit and adaptive approach. In other words, the equations of stellar structure, composition and mesh are solved together, at each timestep and each iteration, and variables are from the current timestep. Within the scheme of an adaptive mesh, mass loss is simply included by adapting an outer boundary condition. More importantly, we find that such an approach is free from some of the notorious numerical instabilities that accompany diffusion calculations.  Such instabilities occur because atomic diffusion, and in particular radiative levitation, can happen on a much shorter time-scale than the evolution of the stellar structure, at least in the outermost layers. In non-simultaneous approaches, where the composition is solved separately from the structure, this can be dealt with by allowing multiple composition timesteps within one evolution timestep (see e.g.~\citealt{turcotte1998}). In addition, evolution timesteps should remain short if diffusion velocities are kept constant during a timestep. This requires lengthy calculations because the radiative accelerations are computationally expensive to evaluate. 

In our calculations, the timesteps are short ($10^2$\,yr) at the start of the evolution but can grow to $10^5$ yr, which is of the same order as for models without diffusion. The timestep size is determined by the change in variables from one timestep to the next. Normally, the temperature and degeneracy have the largest influence but if diffusion acts rapidly the abundance variations limit the timestep size. One might wonder whether the results are sensitive to the timestep size, particularly if mass loss is included. We checked, for one simulation with mass loss, that the results do not change noticeably if timesteps are kept below $10^3$\,yr.

Unfortunately, our numerical scheme requires some effort as well. Like most variables, the radiative accelerations are calculated implicitly; they are not kept constant during a timestep but can change from one iteration to the next. In order to find the correct stellar model, we need to know how a change in $g_{\rm{rad}}$ affects the stellar structure. In other words, we need the derivatives of each $g_{\rm{rad}}$ with respect to the variables that represent the stellar model. STARS normally computes the derivatives numerically but this gives instabilities in the case of $g_{\rm{rad}}$, probably because of their irregularity. Hence their derivatives (or at least some of them) must be obtained analytically. This is done by the OP routines with some modifications, see {Sect.}~\ref{OP}. 

\subsection{Radiative accelerations}\label{OP}
We compute radiative accelerations $g_{\rm{ rad}}$ and Rosseland mean opacities $\kappa_{\rm{R}}$ using OP atomic data. The OP project also provides a set of codes called OPserver  \citep{seaton2005, mendoza2007} which we have rewritten to suit our specific needs. 
The most relevant modifications are summarized here.

\subsubsection{Radiative accelerations in stellar interiors}
For Rosseland mean optical depths $\tau_{\rm R}\geq 1$, the radiative acceleration for element $k$ can be expressed as follows
\begin{equation}\label{gradeq}
g_{{\rm rad}, k} = \frac{\mu\kappa_{\rm{R}}}{\mu_k c}\frac{l}{4\pi r^2}\gamma_k,
\end{equation}
where $\mu$ is the mean atomic weight, $\mu_k$ is the atomic weight of element $k$, $c$ is the speed of light, {$\kappa_{\rm R}$ is the Rosseland mean opacity,} $l$ is the luminosity and $r$ is the radius. The dimensionless parameter $\gamma_k$ depends on the monochromatic opacity data by
\[
\gamma_k=\int \frac{(\sigma_k(u)[1-\exp(-u)]-a_k(u)){\rm d}u}{\sum_k f_k\sigma_k(u)},
\]
where $\sigma_k$ is the cross-section for absorption or scattering of radiation by element $k$, $a(k)$ corrects for electron scattering and momentum transfer to electrons and $f_k$ is the number fraction. 
Eq.~(\ref{gradeq}) is equivalent to equation (6) of \citet{seaton2005} but makes use of the luminosity rather than the effective temperature and stellar radius. The latter two surface quantities are not known during the iterations because the evolution code solves the stellar structure from the centre to the surface.   

\subsubsection{Radiative accelerations for multiple elements}
 For each call OPserver allows the computation of $g_{\rm{rad}}$ for one selected element in the mixture and for multiple mixtures with varying abundances of $k$. Because we are interested in the diffusion of nine elements it saves CPU time to compute $g_{\rm{rad}}$s for multiple elements but only for one mixture in each call.  
 
\subsubsection{Mean ion charges}
 As mentioned before, elements are treated as if they have an mean ion charge $\bar{Z}$. It has been said by \citet{michaud2008} that a disadvantage of OP is that the database does not contain mean charges which are needed to determine the diffusion velocities. We found, however, that $\bar{Z}$ can easily be calculated from their data. That should not be too surprising because the level populations of the ionization states are also needed to compute $g_{\rm{ rad}}$. 

\subsubsection{Interpolation method}\label{interpolationmethod}
 To obtain $g_{\rm{ rad}}$ and $\kappa_{\rm{R}}$ for a specified temperature $T$ and density $\rho$ (converted to electron density $N_e$) OPserver interpolates within a tabulated mesh of  ($T, N_e$) values. OPserver uses two different interpolation schemes. The first is a refined bi-cubic spline interpolation \citep{seaton1993} within the entire ($T,N_e$) mesh. This scheme is applicable for a single chemical mixture only. Secondly, when the chemical abundances vary with stellar depth, a less refined bi-cubic interpolation occurs within a two-dimensional array of $(4,4)$ points around the specified $(T,\rho)$ values. 
It should be clear that the second scheme applies if we want to take into account the variation of abundances owing to atomic diffusion. Unfortunately, this scheme is too crude for pulsation studies for which it is required to have smooth opacity derivatives
\[
 \kappa_{T}=\frac{\partial \log \kappa_{\rm R}}{\partial \log T} \qquad\textrm{and}\qquad
 \kappa_{\rho}=\frac{\partial \log \kappa_{\rm R}}{\partial \log \rho}.
 \]
Therefore, we use the general idea of the second scheme but replace the interpolation method by that developed by \citet{dupret2003} for the purpose of non-adiabatic pulsation computations. This is a local interpolation method by a polynomial of degree 3. It ensures continuity of  $\kappa_{T}$ and  $\kappa_{\rho}$ and their derivatives. Note that, in our scheme, interpolation is made within an array of dimension $(6,6)$ rather than $(4,4)$. This increases  CPU time, so we only use the third interpolation scheme for the pulsation calculations. For the evolution calculations OPserver's second scheme suffices. To demonstrate the differences between the three interpolation methods,  we perform test runs on the same stellar model as used by \citet{seaton2005}. While $\kappa_{\rm R}$ (not shown) itself is not visibly affected by the interpolation method, its derivatives (shown in Fig.~\ref{interpolation}) are.

\subsubsection{Computation of derivatives}\label{derivatives}
As explained in {Sect.~}\ref {evcode}, analytical derivatives of each $g_{\rm{rad}}$ are needed in order for the evolution code to converge towards the correct stellar model. It improves the numerical stability to do this for $\kappa_{\rm{R}}$ and $\bar{Z_i}$ as well. The derivatives we compute are 
\begin{eqnarray}
\frac{\partial\log g_{{\rm rad},i }}{\partial\log T},
 \frac{\partial\log g_{{\rm rad},i}}{\partial\log \rho},
 \frac{\partial\log g_{{\rm rad},i}}{\partial X_i},&\nonumber\\
\frac{\partial \log \kappa_{\rm R}}{\partial \log T} ,
\frac{\partial \log \kappa_{\rm R}}{\partial \log \rho},
\frac{\partial\kappa_{\rm{R}}}{\partial X_i},&\qquad\textrm{for}\quad i=1,...,9\nonumber\\
\frac{\partial\bar{Z}_i}{\partial\log T}\textrm{ and }
 \frac{\partial\bar{Z}_i}{\partial\log \rho}.&\nonumber
\end{eqnarray}

The opacity derivatives are already output by OPserver (see Sect.~\ref{interpolationmethod}). The others we added. Strictly speaking, the derivatives with respect to $\log\rho$ and $\log T$ are not analytic. They are however obtained from the interpolaton curve and so have an analytic form. 

%______________________________________________ 
   \begin{figure}
   \begin{center}
  \includegraphics[angle=-90, width=9cm]{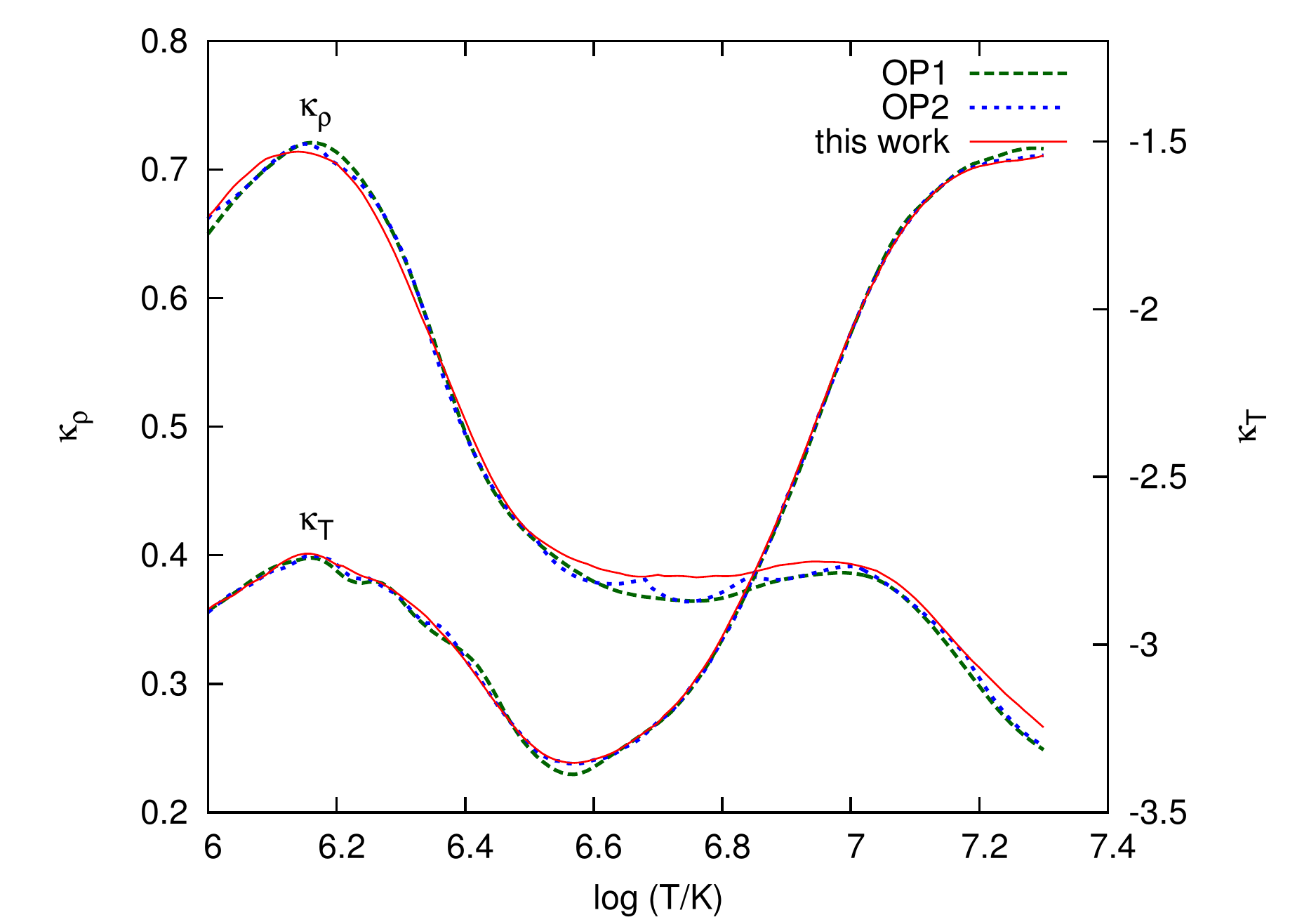}
   \caption{The opacity derivatives $\kappa_{\rho}$ (\emph{left} y-axis) and  $\kappa_{T}$ (\emph{right} y-axis) throughout the star as a function of temperature. OP1 and OP2 indicate OPserver's first and second interpolation scheme, respectively. 
}
              \label{interpolation}
              \end{center}
    \end{figure}
%______________________________________________ 

\subsection{The diffusion equations}\label{burgers}
Having obtained the radiative accelerations $g_{\rm rad}$, we put them in Burgers' diffusion equations \citep{burgers1969},
\begin{eqnarray}\label{4difeq}
\frac{\textrm{d}p_i}{\textrm{d}r}+\rho_i(g-g_{{\rm rad}, i}) -n_i\bar{Z}_ieE=  \nonumber \\
\sum_{j\neq i}^N K_{ij}(w_j-w_i)+\sum_{j\neq i}^{N}K_{ij}z_{ij}\frac{m_jr_i-m_ir_j}{m_i+m_j},
\end{eqnarray}
including the heat flow equations,
\begin{eqnarray}\label{4heat}
\frac{5}{2}n_i k_{\rm B}\nabla T =\frac{5}{2}\sum_{j\neq i}^Nz_{ij}\frac{m_j}{m_i+m_j}(w_j-w_i)-\frac{2}{5}K_{ii}z_{ii}''r_i \nonumber\\
-\sum_{j\neq i}^N\frac{K_{ij}}{(m_i+m_j)^2}(3m_i^2+m_j^2z_{ij}'+0.8m_im_jz_{ij}'')r_i\nonumber\\
+\sum_{j\neq i}^N\frac{K_{ij}m_im_j}{(m_i+m_j)^2}(3+z_{ij}'-0.8z_{ij}'')r_j.
\end{eqnarray}
In addition, we have two constraints, current neutrality,
\begin{equation}\label{4current}
\sum_i \bar{Z}_i n_i w_i = 0
\end{equation}
and local mass conservation,
\begin{equation}\label{4mass}
\sum_i m_i n_i w_i = 0.
\end{equation}
In the above $2N+2$ equations the quantities $p_i$, $\rho_i$, $n_i$, $\bar{Z}_i$ and $m_i$ are the partial pressure, mass density, number density, mean charge and mass for species $i$, respectively. The total number of species (including electrons) is $N$. The $2N+2$ unknown variables are the $N$ diffusion velocities $w_i$, the $N$ heat fluxes $r_i$, the gravitational acceleration $g$ and the electric field $E$. The resistance coefficients $K_{ij}$, $z_{ij}$, $z_{ij}'$ and $z_{ij}''$ are taken from \citet{paquette1986}.  We solve Eqs~(\ref{4difeq})--(\ref{4mass}) with an adapted version of the routine by \citet{thoul1994}. In Appendix \ref{appendix}, we describe how we modify Thoul et al.'s routine to include radiative forces and Paquette et al.'s resistance coefficients. The output of the routine are the diffusion velocities $w_i$. These are then inserted as an advection term in the diffusion equation governing the time evolution of abundance $X_i$,
\[
\frac{\partial X_i}{\partial t} + R_{i}= \frac{1}{\rho r^2}\frac{\partial}{\partial r}\Big(
(D+D_{\rm{T}})\rho r^2\frac{\partial X_i}{\partial r}\Big) - \frac{1}{\rho r^2}\frac{\partial}{\partial r}(\rho r^2 X_i w_i),
\]
where $ R_{i}$ is the change owing to nuclear reactions, $D$ is the combined diffusion coefficient for convection, overshooting and semi-convection \citep{eggleton1972} and $D_{\rm{T}}$ is the diffusion coefficient for turbulent mixing. We take $D_{\rm{T}}$ from \citet{michaud2011},
\begin{equation}\label{turbmodel}
D_{\rm T}=C\,D({\rm He})_0\big( \frac{\rho_0}{\rho}\big)^n, 
\end{equation}
where 
\[
D({\rm He})_0=\Big[\frac{3.3\times 10^{-15}T^{2.5}}{4\rho\ln(1+1.125\times10^{-16}T^3\rho)}\Big]_0
\]
is an approximation to the He diffusion coefficient at a certain reference depth. The subscript 0 indicates the reference depth in terms of the outer mass coordinate $\Delta M=M_*-M_r$. The assumed $D_{\rm T}$ is $C$ times the He diffusion coefficient at $\Delta M_0$ and varying as $\rho^{-n}$. \citet{michaud2011} used a turbulence model with $C=10^4$, $n=4$ and $\Delta M_0=10^{-7.5}\,{\rm M}_{\odot}$ but we shall vary these parameters to examine the influence of the efficiency of turbulent mixing on the model (see Sect.~\ref{turbulence}).

\section{Results}\label{results}
We investigate the effects of atomic diffusion, mass loss, and turbulence in a typical sdB star with total mass $M_*=0.46\,{\rm M}_{\odot}$ and envelope mass $M_{\rm env}=3.5\times10^{-4}\,{\rm M}_{\odot}$. $M_{\rm{env}}$ is the amount of mass above the core boundary and is defined at ZAEHB only, since the H profile can change with time. At the core boundary $X({\rm H})$ starts to increase from 0 to 0.7 at the surface. The initial H profile is fitted to the profile on the RGB according to \citet{hu2009}. The ZAEHB model has a $T_{\rm eff}=30\,{\rm kK}$ and $\log (g/{\rm cm\,s}^{-2})=5.8$, and it has a metallicity at all depths of $Z=0.02$ with metal mixture of \citet{grevesse1993}. Although sdB stars could have a range of original metallicities, \citet{michaud2011}'s diffusion calculations showed that the surface abundances are poorly sensitive to the initial metallicity. Thus we do not expect our results to be much affected by the choice of initial abundances. 

Starting at the ZAEHB we compute the stellar evolution until the end of core He burning under different assumptions for mass loss and turbulent mixing. Thereafter, we compute non-adiabatic oscillations of the sdB models. We restrict ourselves to modes of spherical degree $\ell\leq2$, because higher  $\ell$ values are geometrically disfavoured for observation. We search for eigenfrequencies in the range of $0.1\leq\omega\leq20$ where $\omega$ is the dimensionless angular eigenfrequency, $\omega=2\pi f \tau_{\rm{dyn}}$, $f$ is the frequency and $\tau_{\rm dyn}$ is the dynamical time-scale.

\subsection{Atomic diffusion, no mass loss/turbulence}\label{diffusion}
In Fig.~\ref{grad} we show radiative accelerations for C, N, O, Ne, Mg, Fe and Ni in the ZAEHB model. H and He are not shown because their $g_{\rm{rad}}$ are so low that they lie outside the range of the plot. Because gravitational settling and radiative levitation are the dominant diffusion processes, gravity is also plotted for comparison. 

Fig.~\ref{grad} tells us that Fe and Ni are levitated, whereas the other elements sink sooner or later. This is confirmed in Fig.~\ref{abund} which illustrates the time evolution of the interior abundances from ZAEHB to $10^8$ yr. Notice that Ni accumulates to mass fractions comparable to (or even higher than) Fe in the outer layer, while it starts out with a much smaller mass fraction. This implies that Ni is (at least) as important for mode driving as Fe. Notice also the development of an iron-group convection zone within $10^4$ yr around $\Delta M=10^{-10}\,{\rm M}_{\odot}$. The  abundance profiles in this region are flattened and do not have a detailed shape as predicted by equilibrium diffusion theory.

%______________________________________________ 
   \begin{figure}
   \begin{center}
  \includegraphics[angle=-90, width=9cm]{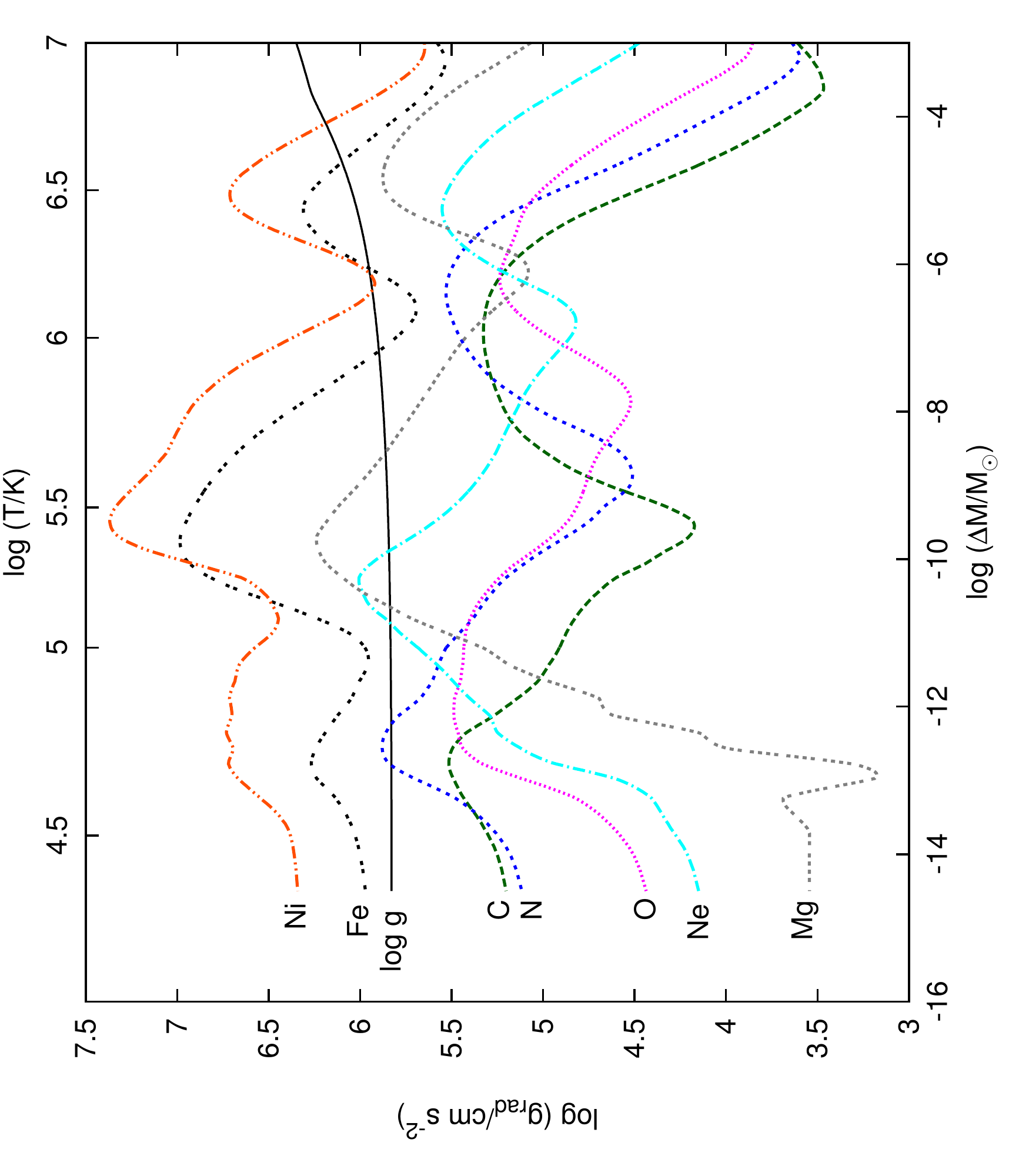}
   \caption{Radiative accelerations in the stellar envelope as a function of mass coordinate (\emph{bottom} axis) and temperature (\emph{top} axis). Gravitational acceleration is represented by the solid line. This model is at the ZAEHB with $T_{\rm eff}=30,000\,{\rm K}$ and $\log (g/{\rm cm\,s}^{-2})=5.8$ at the surface. 
   }
              \label{grad}
              \end{center}
    \end{figure}
%______________________________________________ 
   \begin{figure*}
   \begin{center}
  \includegraphics[angle=-90, width=18cm]{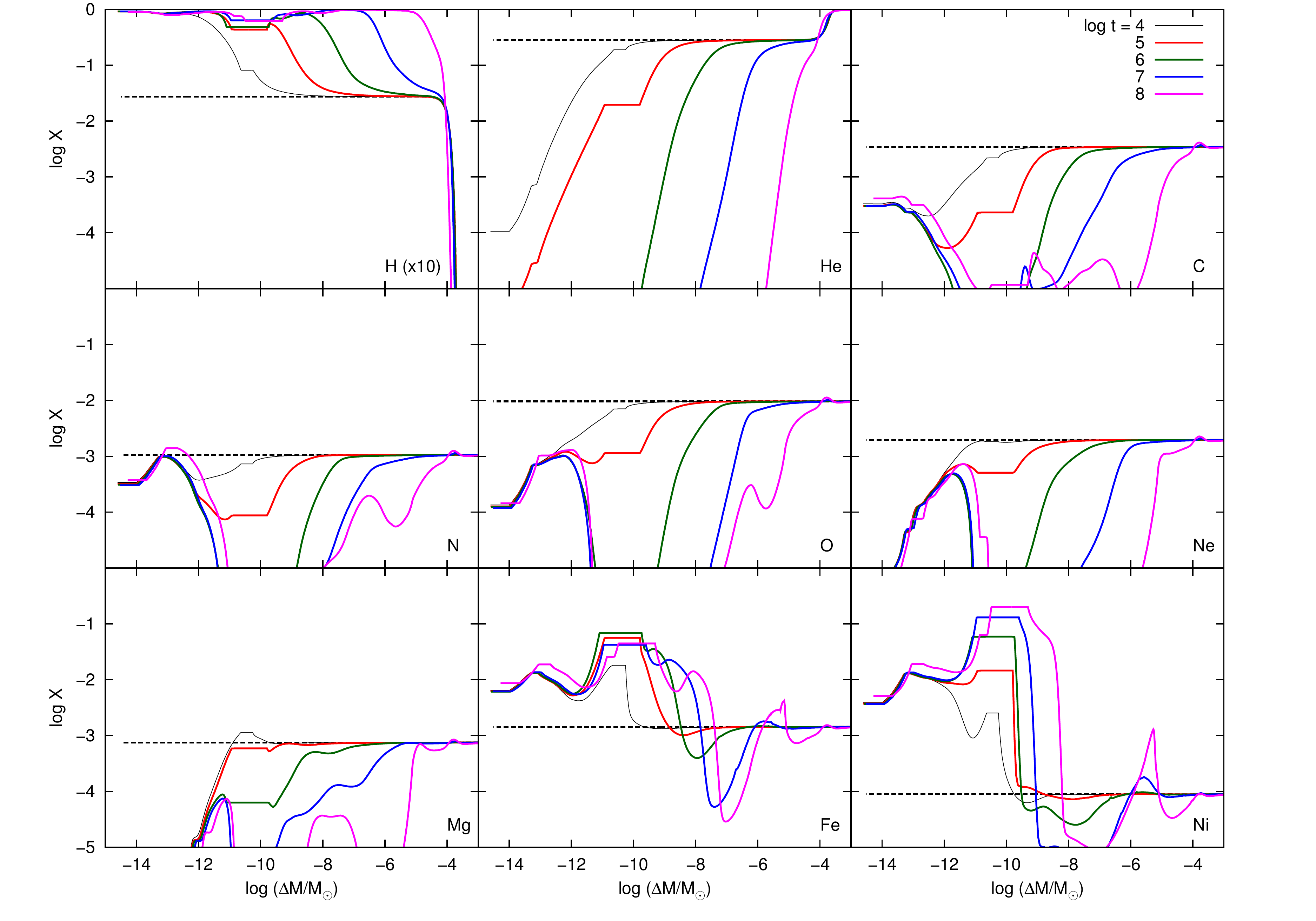}
   \caption{Interior abundance profiles at different stellar ages. The horizontal dashed lines give initial abundances at ZAEHB, and the solid curves are at EHB ages $\log (t/{\rm yr})=4$, 5, 6, 7 and 8 as indicated in the legend. Note that in the case of hydrogen we plot $10\log X({\rm H})$ for visibility. These simulations are with neither mass loss nor turbulence. }
              \label{abund}
              \end{center}
    \end{figure*}
%______________________________________________ 
   \begin{figure}
   \begin{center}
  \includegraphics[angle=-90, width=8cm]{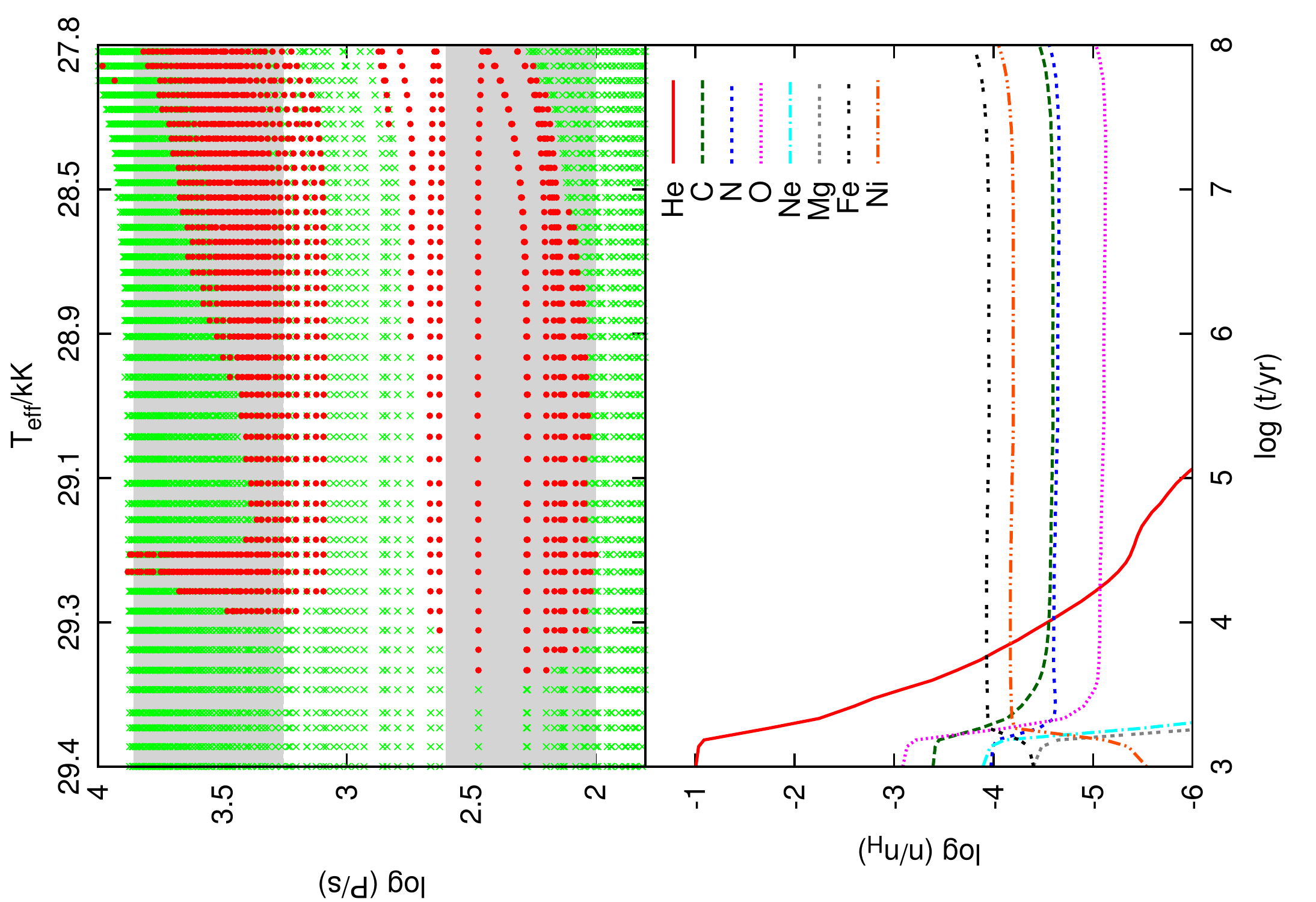}
   \caption{ {An evolutionary sequence of a model with neither mass loss nor turbulence. \emph{Top} panel: time evolution of pulsation periods. Red dots indicate unstable pulsation modes while green crosses are stable modes. The shaded areas are the period ranges for the observed $p$- (lower) and $g$-modes (upper). \emph{Bottom} panel: time evolution of surface abundances. The \emph{top} x-axis displays the $T_{\rm eff}$ corresponding to the stellar age on the \emph{bottom} x-axis.}
     }
              \label{mode1}
              \end{center}
    \end{figure}
%______________________________________________ 

In the upper panel of Fig.~\ref{mode1}, we show the computed pulsation periods as a function of stellar age (bottom axis) and $T_{\rm eff}$ (top axis).  {We also plot the ranges of observed periodicities.} We see that this particular simulation represents a hybrid pulsator, showing both unstable $p$- and $g$-modes. {The total period range of unstable modes matches the observed range (100\,s -- 2\,hr) well. In the transition region between the $p$- and $g$-modes (400\,s -- 30\,min) the agreement is not as good, but unconfirmed low-amplitude peaks have been observed in that region (\citealt{baran2011} and references therein).} 

It is important to realize that the low-degree ($\ell=1,2$) $g$-modes would not be driven at such high $T_{\rm eff}\approx29\,{\rm kK}$ without Ni diffusion. We find that Ni plays an even bigger role in mode driving than first suggested by \citet{jeffery2006}. In their models only $\ell>2$ modes, which are unlikely to be observed, are driven at  such high $T_{\rm{eff}}$s. This is because Ni accumulates more than Fe in the driving region which was not realized before without diffusion calculations. We shall present a detailed investigation of the sdB instability strips elsewhere (Hu et al.~in prep.).
 
The lower panel of Fig.~\ref{mode1} shows the time evolution of the surface abundances. The problem of the He abundance is clearly illustrated because $n_{\rm{He}}/n_{\rm{H}}$ decreases to below $10^{-4}$ within $10^4$ yr, or 0.01 per cent of the EHB lifetime, whereas the average observed  value is around $10^{-2}$ (see e.g.~\citealt{edelmann2003}). The other surface abundances are qualitatively in agreement with observations. C, N, O and Fe are within the observed ranges given by \citet{geier2010}. Ne and Mg are almost completely depleted in our models but this could be consistent with measurements that give only upper limits. Ni is 1.5 dex above solar. This is on the high side but still in agreement with observations that go up to $1-2$ dex above solar \citep{otoole2006}.  For the remainder of this work, we shall discuss only the abundances of He, Fe and Ni, since only these are relevant for our argumentation for constraining the mass loss and turbulent mixing.
 
\subsection{Atomic diffusion and mass loss}\label{massloss}
The fact that we observe sdB pulsations puts an upper limit on the mass-loss rates, at least for the pulsators. After all, Fig.~\ref{mode1} shows that it takes over $10^4$ yr before enough iron-group elements have accumulated to drive pulsations. The accumulation at this age goes down to $\Delta M\approx10^{-10}\,{\rm M}_{\odot}$ in our models (see Fig.~\ref{abund}). So if this outer mass is removed within $10^4$ yr, or equivalently $\dot{M}\gtrsim 10^{-14}\,{\rm M}_{\odot}\,{\rm yr}^{-1}$, then there is no chance for the metals to build up.

To verify this estimate we perform simulations with $\dot{M}=10^{-15}$, $5\times10^{-15}$, $10^{-14}$ and $10^{-13}\,{\rm M}_{\odot}\,{\rm yr}^{-1}$. We assume that the mass loss is constant and chemical homogeneous. The possibilities of a variable and selective mass loss are discussed later. We found that numerical instabilities occur for models that develop an iron-group convection zone and experience mass loss at the same time (see also \citealt{charbonneau1993}). This happens for $\dot{M}=10^{-15}\,{\rm M}_{\odot}\,{\rm yr}^{-1}$ in our models. To reduce instabilities we assume, only in this case, overshooting of the iron-group convection zone to the atmosphere. The extra mixing occurs only in the outer $10^{-11}\,{\rm M}_{\odot}$ and  has a negligible effect. Still, this simulation is terminated after $10^7$ yr because of poor convergence and hence long computing time.   

In the upper panels of Fig.~\ref{mode2} we show the effect of these mass-loss rates on the stability of the pulsation modes. We find that, as expected, for $\dot{M}\geq10^{-14}\,{\rm M}_{\odot}\,{\rm yr}^{-1}$ pulsations cannot be driven. For $\dot{M}=5\times10^{-15}\,{\rm M}_{\odot}\,{\rm yr}^{-1}$ pulsations are driven only during the first $10^5$\,yr of evolution. This time-scale is $100\times$ smaller than that found by \citet{fontaine2006} for a comparable mass-loss rate. It is too short compared to the evolutionary timescale of $10^8$ yr to explain the fraction of sdBs that are variable (about 10 per cent) as suggested by \citet{fontaine2006}. The difference is caused by their assumption that an equilibrium between gravitational settling and radiative levitation can be reached before the onset of mass loss. Furthermore, they did not include diffusion during the evolution while mass is lost. In our models, however, diffusion operates at the same time as mass loss and both processes start at the ZAHB.  We find that, under these circumstances, the rates should be $\dot{M}\leq10^{-15}\,{\rm M}_{\odot}\,{\rm yr}^{-1}$ if pulsations are to be driven for a significant amount of time.

Remember that, if mass loss were to explain the observed He abundances, then the rates should be in the range $10^{-14}\lesssim\dot{M}/{\rm M}_{\odot}\,{\rm yr}^{-1}\lesssim10^{-13}$ \citep{fontaine1997,unglaub2001}. This can also be seen in Fig.~\ref{mode2} (lower panels) where the surface abundances of He, Fe and Ni are plotted. Our results show that these mass-loss rates are not consistent with the observed pulsations in sdB stars. One could argue that the constant sdBs are experiencing mass loss while the pulsators are not (or much less). But then there should be a negative correlation between the He abundance and the presence of pulsations. This has not been observed despite many efforts (e.g.~\citealt{otoole2006, blanchette2008}). 

Another possibility is a variable mass loss during stellar evolution. When the mass-loss rate is low, diffusion can build up enough Fe and Ni to drive pulsations. When the rate increases, the Fe and Ni reservoir is emptied and the star becomes constant. We use the mass loss recipe by \citet{vink2002} to estimate how much the mass-loss rate could vary. Applying their equation (5) to our models, we find that the rate increases gradually from $\dot{M}\approx4\times 10^{-13}$ to $2\times 10^{-12}\,{\rm M}_{\odot}\,{\rm yr}^{-1}$ during the sdB lifetime. Although the rates themselves could be overestimated due to an underestimation of the terminal velocity \citep{unglaub2008}, the dependence on the stellar parameters {($T_{\rm eff}, L_*, M_*, Z_*$)} should hold \citep{vink2000}. Thus we expect the increase in the mass-loss rate, which is mostly due to the increasing luminosity as the star evolves\symbolfootnote[1]{{The dependence of $\dot{M}$ on $Z_*$ has a small negative effect during the evolution because the metallicity decreases as the lighter metals sink. Although one might expect that the mass-loss rate increases with [Fe/H], one should not forget that the relative role of Fe, compared to lighter elements, diminishes for weaker winds \citep{vink2001}. A detailed study that evaluates the contribution of different metals to the mass loss is needed but beyond the scope of this work.}}, to remain {approximately} valid. This factor 5 over the sdB lifetime is not sufficient to obtain the rates needed to keep He from settling ($\dot{M}>10^{-14}\,{\rm M}_{\odot}\,{\rm yr}^{-1}$) if starting with a rate low enough to drive pulsations ($\dot{M}\leq 10^{-15}\,{\rm M}_{\odot}\,{\rm yr}^{-1}$).  Hence we find it unlikely that mass loss is the dominant process for slowing down atomic diffusion in sdB stars{, although a more sophisticated understanding of mass loss is required to draw any definite conclusions.}

Fig.~\ref{mode2} also shows that, in the presence of mass loss, the surface abundances are poor indicators of whether a star is pulsating. For example, during the first $10^6$\,yr of the simulation with $\dot{M}=10^{-14}\,{\rm M}_{\odot}\,{\rm yr}^{-1}$, both Fe and Ni reach overabundances above 1 dex while no modes are excited. This is because, although the surface is enriched, Fe en Ni are actually depleted in the driving region.\symbolfootnote[2]{The driving is caused by the iron-group opacity bump at $\log (T/{\rm K})\approx 5.3$. This corresponds to $\Delta M_0\approx10^{-10}\,{\rm M}_{\odot}$ in our stellar models.} In Fig.~\ref{abund2} we plot the time evolution of the interior abundances of Fe and Ni. Basically what happens is that, as the outermost  layers of the star are removed, the regions underneath become exposed. Thus as time progresses and more mass is peeled away, the surface abundances are determined by processes that occurred deeper in the star. After $10^5$ yr the surface is enriched in Fe and Ni because the region where accumulation took place earlier ($\Delta M\approx\dot{M}\Delta t= 10^{-9}\,{\rm M}_{\odot}$) is now exposed. However, the driving region at this time corresponds to a region that was previously deeper inside the star and from which Fe and Ni have been transported away. After $10^7$ yr so much mass is stripped away that the depleted regions are now at the surface, causing underabundances of Fe and Ni. The actual situation is more complicated because diffusion operates at the same time as mass is lost and the diffusion velocities change as the star evolves.
%______________________________________________ 
   \begin{figure*}
   \begin{center}
  \includegraphics[angle=-90, width=18cm]{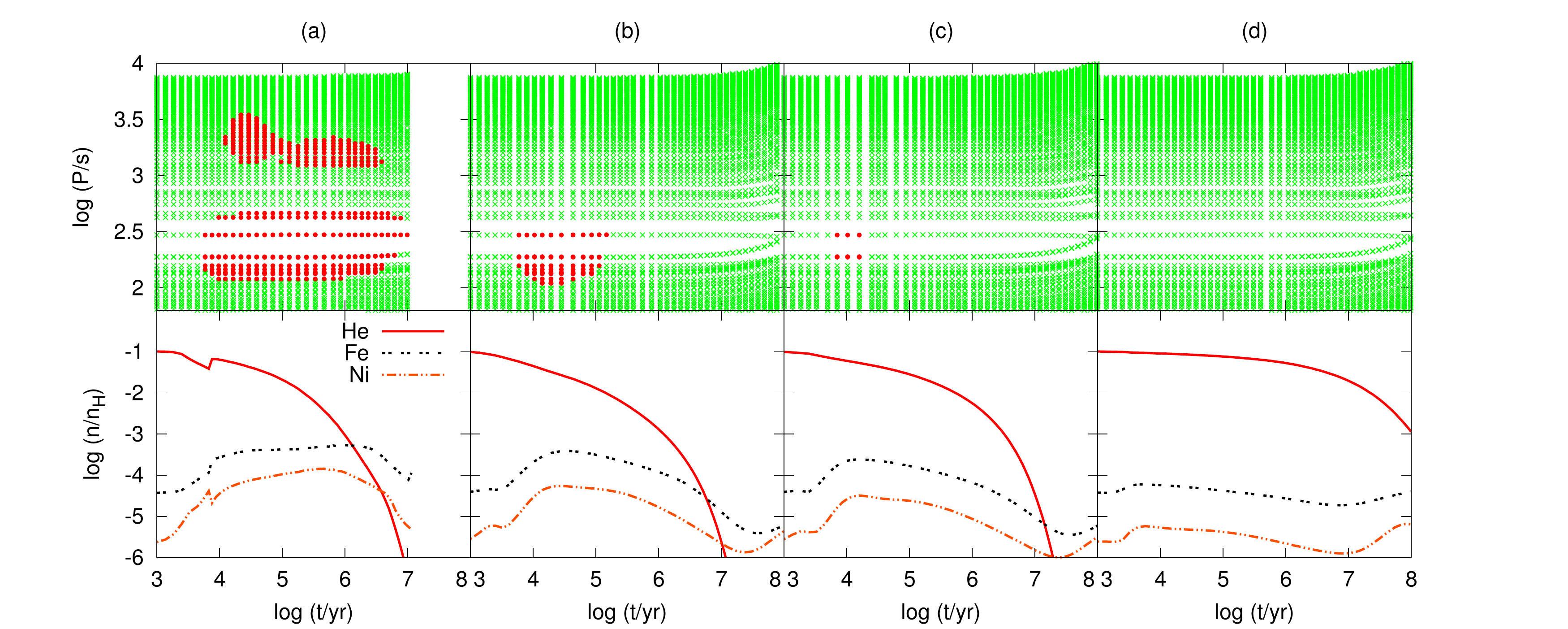}
   \caption{Models with mass loss. \emph{Upper} panels show pulsations as a function of time for mass-loss rates (a) $10^{-15}$, (b) $5\times10^{-15}$, (c) $10^{-14}$ and (d) $10^{-13}\,{\rm M_{\odot}yr^{-1}}$. The red dots are unstable modes while the green crosses are stable. \emph{Lower} panels show surface abundances of He, Fe, and Ni corresponding to these mass-loss rates. {The kink in the surface abundances for $\dot{M}=10^{-15}\,{\rm M}_{\odot}\,{\rm yr}^{-1}$ at $10^4$ yr  is caused by the overshoot of the iron-group convection zone assumed in this particular simulation.}
}
              \label{mode2}
              \end{center}
    \end{figure*}
%______________________________________________ 
   \begin{figure}
   \begin{center}
  \includegraphics[angle=-90, width=9cm]{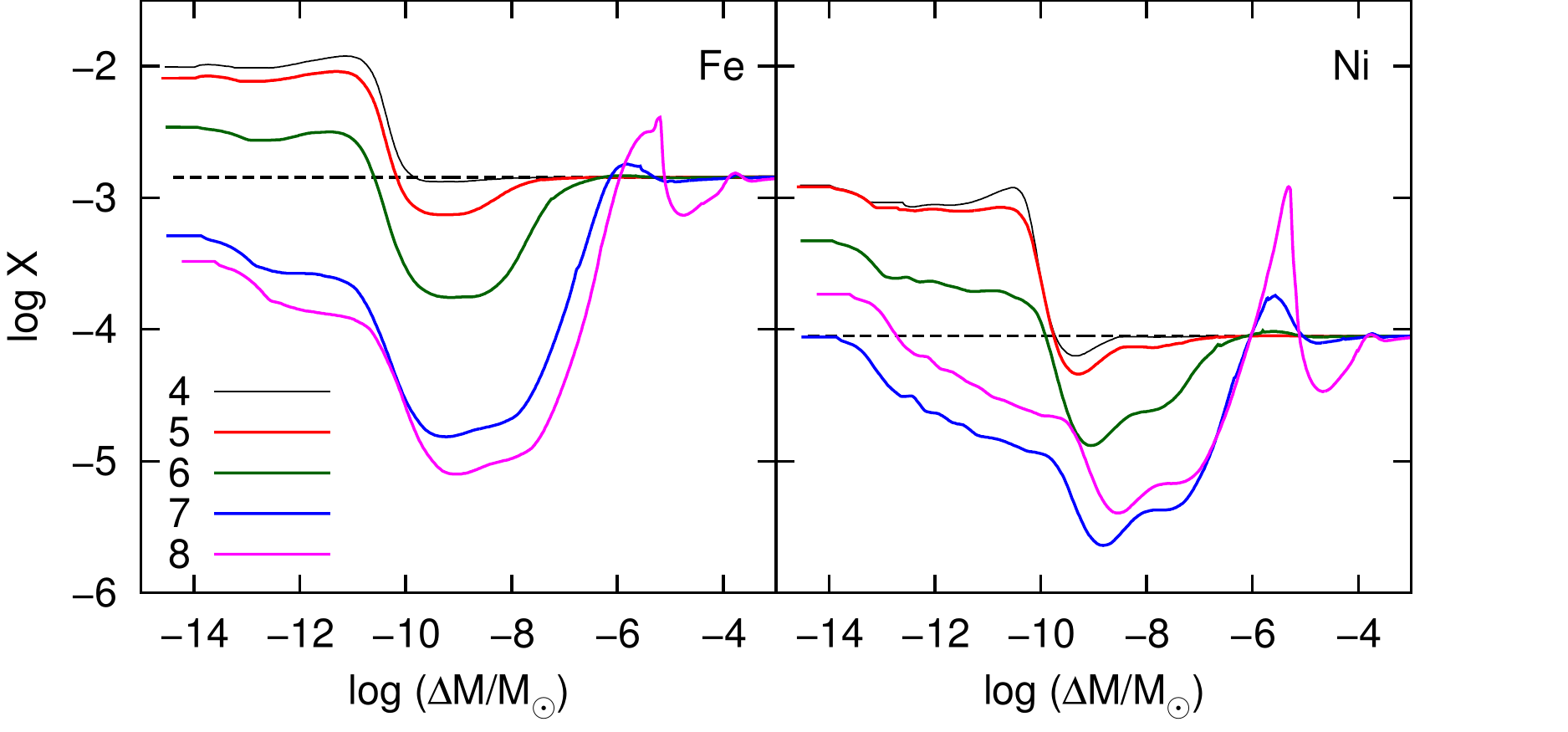}
   \caption{Interior abundance profiles of Fe and Ni at different stellar ages for the simulation with $\dot{M}=10^{-14}\,{\rm M}_{\odot}\,{\rm yr}^{-1}$. Lines are as in Fig.~\ref{abund}. }
              \label{abund2}
              \end{center}
    \end{figure}
%______________________________________________  

\subsection{Atomic diffusion and turbulent mixing}\label{turbulence}
If mass loss is not responsible for retarding atomic diffusion then perhaps turbulence is. \citet{michaud2011} showed that their turbulence model (Eq.~\ref{turbmodel}) with $C=10^4$, $n=4$ and $\Delta M_0=10^{-7.5}$ M$_{\odot}$\symbolfootnote[3]{This results in the outer $\Delta M\approx10^{-7}$ M$_{\odot}$ of the star being completely mixed. The mixed mass varies slightly with atomic species.} can explain most observed abundance anomalies in field and globular cluster sdBs. In their models Fe is near solar throughout the entire mixed region, including the driving region. This raises doubt as to whether pulsations can be excited. 
As they conclude themselves, their work has yet to be tested with asteroseismology. For this purpose, we run simulations with $\Delta M_0=10^{-8.5}$,  $10^{-7.5}$ and $10^{-6.5}\,{\rm M}_{\odot}$, corresponding to efficient mixing of the outer $\Delta M\approx10^{-8}$, $10^{-7}$ and $10^{-6}\,{\rm M}_{\odot}$, respectively. The other parameters are kept at $C=10^4$ and $n=4$.

In {Fig.~\ref{mode3}a--c} we show the effect of these three turbulence models on the stability of the pulsation modes. We see that mixing the outer mass of $\Delta M\approx10^{-8}$ M$_{\odot}$ allows some modes to be driven. Although Fe stays near solar, Ni can still accumulate and facilitate mode excitation. However, the He surface abundance still drops too quickly to explain the observations. Mixing more mass ($\Delta M\approx10^{-7}$ M$_{\odot}$) slows down He settling further but it also prevents the accumulation of both  Fe and Ni and hence mode excitation. Interestingly, if mixing occurs down to $\Delta M\approx10^{-6}$ M$_{\odot}$ a few pulsation modes can be excited after $8\times10^7$ yr. This can be understood from Fig.~\ref{abund} where we see a second, very small accumulation of Fe and Ni appearing around $\Delta M\approx10^{-6}$ M$_{\odot}$) at late stellar ages. However, the number of unstable modes is very small and cannot explain the many (dozens) frequencies observed in sdB stars. 
%______________________________________________ 
   \begin{figure*}
   \begin{center}
  \includegraphics[angle=-90, width=18cm]{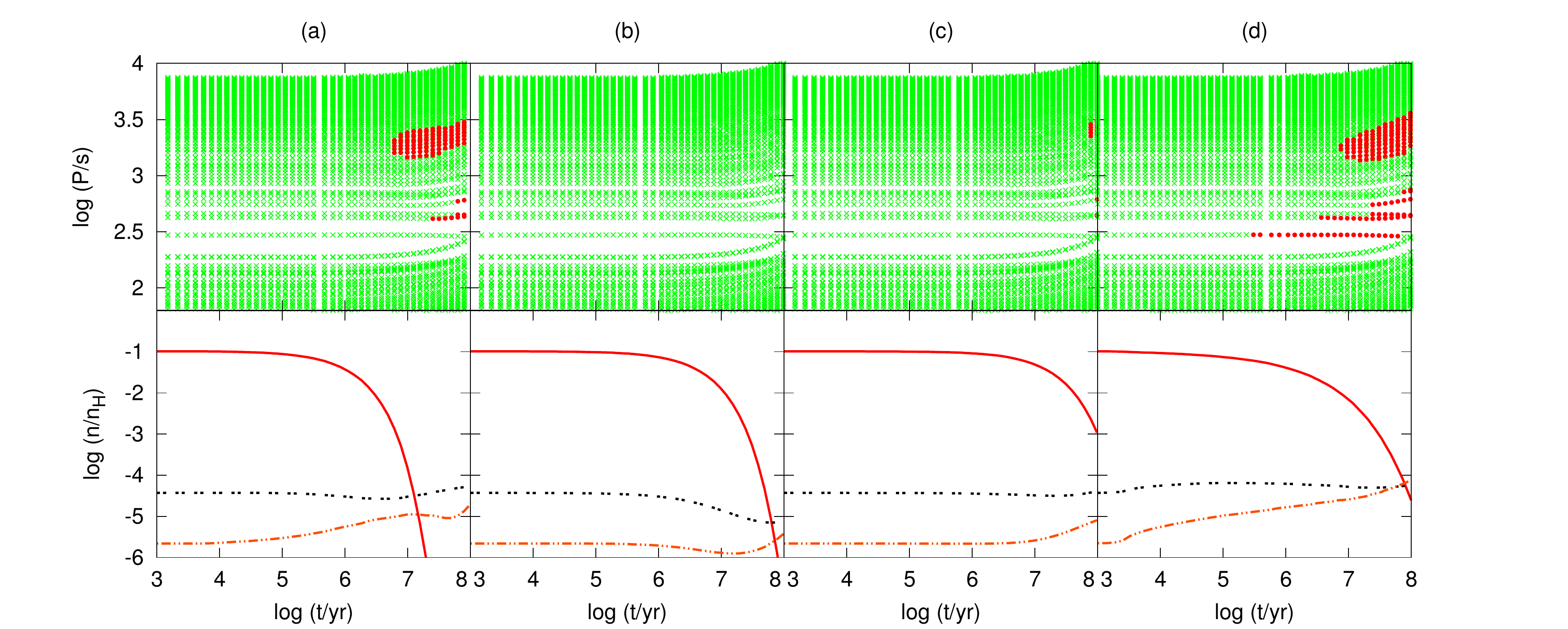}
   \caption{Models with turbulent mixing. \emph{Upper} panels show pulsations as a function of time for turbulence models (a) $\Delta M_0=10^{-8.5}\,{\rm M}_{\odot}$, $C=10^4$, $n=4$ (b) $\Delta M_0=10^{-7.5}\,{\rm M}_{\odot}$, $C=10^4$, $n=4$, (c)   $\Delta M_0=10^{-6.5}\,{\rm M}_{\odot}$, $C=10^4$, $n=4$ and (d) $\Delta M_0=10^{-8.2}\,{\rm M}_{\odot}$, $C=10^2$, $n=1$.  \emph{Lower} panels show surface abundances of He, Fe, and Ni corresponding to these turbulence models. Symbols and lines are as in Fig.~\ref{mode2}.
}
              \label{mode3}
              \end{center}
    \end{figure*}
%______________________________________________ 

It would at first sight seem that turbulence cannot be reconciled with mode excitation either. However, this can be solved by adjusting the efficiency of turbulence because the amount of mixing can vary with atomic species. After all, the diffusion velocities of Fe and Ni  are much larger than for He, so a weak amount of turbulence could retard He settling while having only a small effect on Fe and Ni levitation. This possibility is investigated by varying the parameters $C$, $n$ and $\Delta M_0$ in Eq.~(\ref{turbmodel}). Indeed, we find for $C=100$, $n=1$ and $\Delta M_0=10^{-8.2}\,{\rm M}_{\odot}$ that He settles from $0.1$ to $10^{-4}$ during the sdB lifetime while Fe and Ni can accumulate enough to excite pulsations (see {Fig.~\ref{mode3}d}).
Efficient mixing only occurs in the outer $\Delta M\approx10^{-9}\,{\rm M}_{\odot}$, but  weak mixing occurs up to $\Delta M\approx10^{-6}\,{\rm M}_{\odot}$. This can be seen by the flattening of the abundance profiles plotted in Fig.~\ref{abund3}. Of course, it cannot be guaranteed that the consistency between abundance predictions and observations found by \citet{michaud2011} can be recovered. However, their turbulence model was fine-tuned to keep the Fe abundance near solar, and this remains true in our model. 

%______________________________________________ 
  \begin{figure}
   \begin{center}
  \includegraphics[angle=-90, width=9cm]{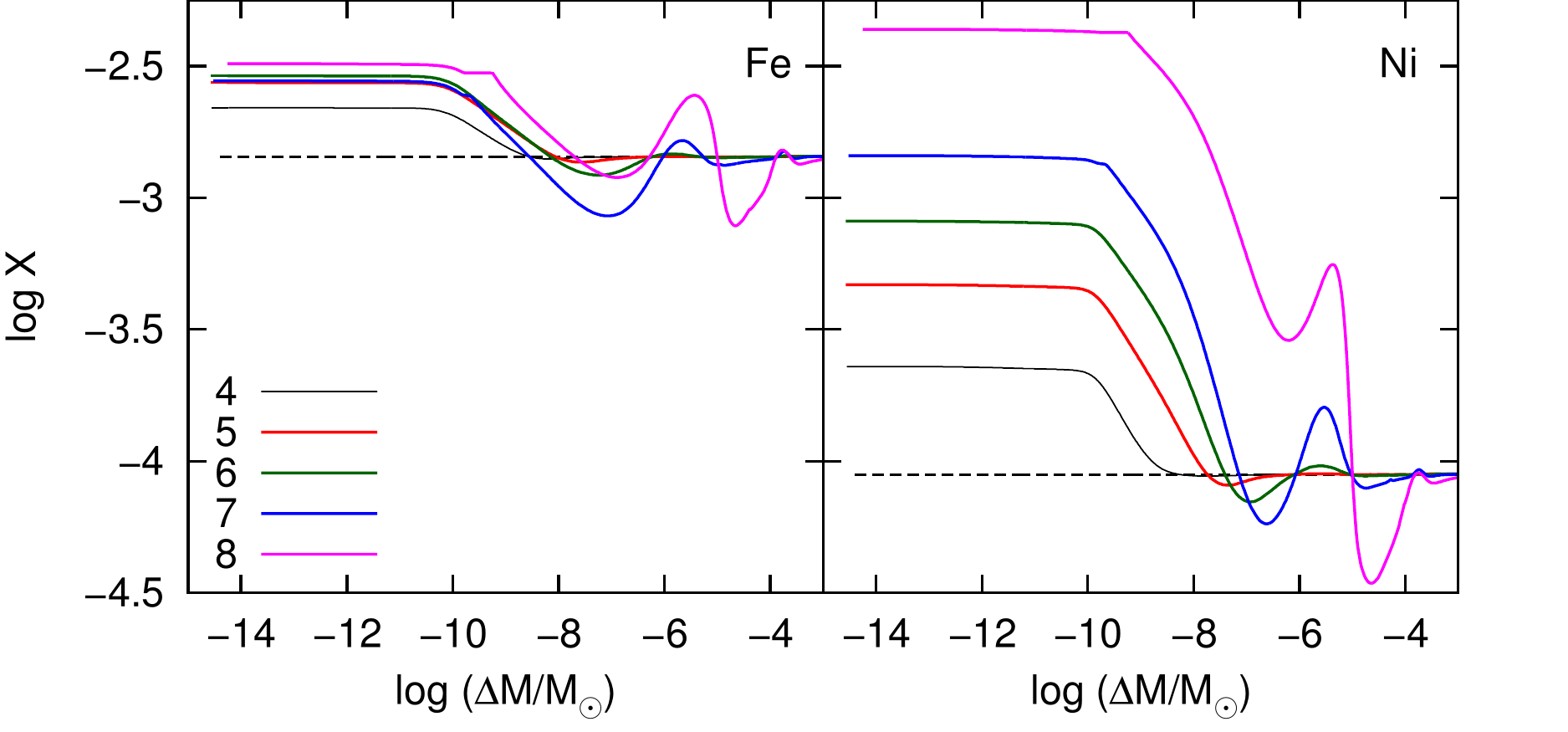}
   \caption{Interior abundance profiles of Fe and Ni at different stellar ages for the simulation with $\Delta M_0=10^{-8.2}\,{\rm M}_{\odot}$, $C=10^2$, $n=1$. Lines are as in Fig.~\ref{abund}. }
              \label{abund3}
              \end{center}
    \end{figure}
%______________________________________________  

{In Table \ref{table} we summarize our results in terms of the fraction of its lifetime the sdB star has (i) He surface abundances in the typically observed range of $10^{-4}-0.1$, (ii) unstable pulsation modes or (iii) both. We see that, of all the scenarios we examined, only the sdB models with mixing in the outer $10^{-6}\,{\rm M}_{\odot}$ spend a significant time fulfilling both conditions.}
 \begin{table*}
\begin{minipage}{126mm}
{
\caption{Percentage of EHB lifetime ($\sim$$10^8$ yr) spent with He surface abundance in the observed range and/or with pulsation modes excited.} \label{table}
\begin{center}
\begin{tabular}{l r r  r r r r r r r }
\hline
Simulation & M0T0 & M1  & M2 & M3 & M4 & T1 & T2 & T3 & T4\\
\hline
(i) He condition & $<1$ & 4 & 5 & 6 & 100 & 7 & 50 &100  &65 \\
(ii) pulsation condition & $>99$ &$\gtrsim 10$& $<1$ & $<1$ & 0 & 90 & 0 & 40 &$>99$ \\
(iii) both conditions& 0 & 4 & $<1$ & $<1$ & 0  &  5 &  0 & 40 & 64 \\
\hline
\end{tabular}
\end{center}
\medskip
Simulation (M0T0) is with neither mass loss nor turbulence. The simulations with mass loss are (M1) $\dot{M}=10^{-15}\,{\rm M}_{\odot}\,{\rm yr}^{-1}$, (M2) $\dot{M}=5\times10^{-15}\,{\rm M}_{\odot}\,{\rm yr}^{-1}$, (M3) $\dot{M}=10^{-14}\,{\rm M}_{\odot}\,{\rm yr}^{-1}$ and (M4) $\dot{M}=10^{-13}\,{\rm M}_{\odot}\,{\rm yr}^{-1}$. The simulations with turbulence are (T1) $\Delta M_0=10^{-8.5}\,{\rm M}_{\odot}$, $C=10^4$, $n=4$, (T2) $\Delta M_0=10^{-7.5}\,{\rm M}_{\odot}$, $C=10^4$, $n=4$, (T3) $\Delta M_0=10^{-6.5}\,{\rm M}_{\odot}$, $C=10^4$, $n=4$ and (T4) $\Delta M_0=10^{-8.2}\,{\rm M}_{\odot}$, $C=10^2$, $n=1$. The numbers give the percentage of its life the sdB star spends with (i) He surface abundance in the range $10^{-4}-0.1$, (ii) unstable pulsation modes or (iii) both conditions fulfilled.  Note that simulation (M1) was terminated after $10^7$ yr owing to poor convergence, so we only provide an estimate here. 
}
\end{minipage} 
\end{table*}

\section{Summary and discussion}\label{summary}
We have constructed sdB evolutionary and seismic models with atomic diffusion, including the often neglected radiative forces. Our aim is to discriminate between mass loss and turbulence in sdB stars using non-adiabatic asteroseismology. We performed a stability analysis on stellar models with various mass-loss rates and turbulence models. We found that the mass-loss rates required to match the observed He abundances are not consistent with observed pulsations. However, weak turbulent mixing  of the outer $10^{-6}\,{\rm M}_{\odot}$ can also explain the He abundances while still allowing pulsation modes to be driven. The presence of unstable modes is very sensitive to the turbulence model. This could explain why some sdBs are pulsating while others, with similar $T_{\rm eff}$ and $\log g$, are {constant}.

Although our results favour turbulence over mass loss, we should remain cautious because the turbulence is not yet supported by a physical model. Possibly, thermohaline mixing in the presence of a mean molecular weight $\mu$ inversion could play a role \citep{theado2009}. But a detailed investigation including the effect of radiative accelerations on the $\mu$-inversion instability is still missing. Also, the possibility of overshooting beyond the surface convection zones of sdB stars should be examined with convection models that are more realistic than mixing-length theory. Such studies for A stars show that overshoot can cause the regions between surface convection zones to completely mix (\citealt{kupka2002, freytag2004}). Our models based on the mixing-length theory develop a relatively broad iron-group convection zone on a very short time-scale.
Another candidate for causing turbulent mixing could be rotation. Although sdB stars typically have slow surface rotation ($v\sin i<10\,{\rm km\,s}^{-1}$), they might hide a rapidly rotating core, a relic from their previous evolution \citep{kawaler2005}. The differential rotation could cause shear turbulence but this has never been examined in sdB stars as far as we know. 
We intend to study the influence of these additional mixing processes  in a forthcoming paper.

In this study we used observed He abundances to constrain mass-loss rates and turbulent mixing. Metals are not trustworthy in this respect because of the possibility of weak metallic winds ($\dot{M}\lesssim10^{-16}\,{\rm M}_{\odot}\,{\rm yr}^{-1}$) in sdB stars as suggested by \citet{unglaub2008}. Such low mass-loss rates could still act to change the atmospheric metal abundances without interfering with mode driving. This  complicates or even prevents the discrimination between variable and constant sdBs on the basis of atmospheric abundances, in agreement with hitherto unsuccessful attempts \citep{otoole2004,otoole2006,blanchette2008}. 

It is noteworthy that Ni is as important for mode driving as Fe. Indeed, the diffusive equilibrium abundances in the driving region are comparable despite nickel's lower initial abundance. Thus Ni plays an even more important role than previous studies of \citet{jeffery2006} and \citet{hu2009} indicated because there it was assumed that Ni was enhanced by the same factor as Fe. Our models show excitation of low-degree ($\ell=1,2$) $g$-modes at relatively high $T_{\rm eff}$ ($29,000$ K), so it is likely that the blue-edge problem can be resolved. A detailed study of the instability strips is underway (Hu et al.~in preparation).

\section*{Acknowledgments}
We thank J.~J.~Eldridge and S.~Th\'eado for helpful discussions. We also like to thank the referee for useful comments. HH is supported by the Netherlands Organisation for Scientific Research (NWO). 

\bibliographystyle{mn2e}
\bibliography{hu2011}

\begin{thebibliography}{}

\bibitem[\protect\citeauthoryear{{Angulo} \& {et al.}}{{Angulo} \& {et
  al.}}{1999}]{angulo1999}
{Angulo} C.,  {et al.} 1999, Nuclear Physics A, 656, 3

\bibitem[\protect\citeauthoryear{{Badnell}, {Bautista}, {Butler}, {Delahaye},
  {Mendoza}, {Palmeri}, {Zeippen} \& {Seaton}}{{Badnell}
  et~al.}{2005}]{badnell2005}
{Badnell} N.~R.,  {Bautista} M.~A.,  {Butler} K.,  {Delahaye} F.,  {Mendoza}
  C.,  {Palmeri} P.,  {Zeippen} C.~J.,    {Seaton} M.~J.,  2005, \mnras, 360,
  458

\bibitem[\protect\citeauthoryear{{Baran}, {Kawaler}, {Reed}, {Quint},
  {O'Toole}, {{\O}stensen}, {Telting}, {Silvotti}, {Charpinet},
  {Christensen-Dalsgaard}, {Still}, {Hall} \& {Uddin}}{{Baran}
  et~al.}{2011}]{baran2011}
{Baran} A.~S.,  {Kawaler} S.~D.,  {Reed} M.~D.,  {Quint} A.~C.,  {O'Toole}
  S.~J.,  {{\O}stensen} R.~H.,  {Telting} J.~H.,  {Silvotti} R.,  {Charpinet}
  S.,  {Christensen-Dalsgaard} J.,  {Still} M.,  {Hall} J.~R.,    {Uddin} K.,
  2011, \mnras, p.~851

\bibitem[\protect\citeauthoryear{{Blanchette}, {Chayer}, {Wesemael},
  {Fontaine}, {Fontaine}, {Dupuis}, {Kruk} \& {Green}}{{Blanchette}
  et~al.}{2008}]{blanchette2008}
{Blanchette} J.,  {Chayer} P.,  {Wesemael} F.,  {Fontaine} G.,  {Fontaine} M.,
  {Dupuis} J.,  {Kruk} J.~W.,    {Green} E.~M.,  2008, \apj, 678, 1329

\bibitem[\protect\citeauthoryear{{B{\"o}hm-Vitense}}{{B{\"o}hm-Vitense}}{1958}%
]{bohm-vitense1958}
{B{\"o}hm-Vitense} E.,  1958, \zap, 46, 108

\bibitem[\protect\citeauthoryear{{Brown}, {Ferguson}, {Davidsen} \&
  {Dorman}}{{Brown} et~al.}{1997}]{brown1997}
{Brown} T.~M.,  {Ferguson} H.~C.,  {Davidsen} A.~F.,    {Dorman} B.,  1997,
  \apj, 482, 685

\bibitem[\protect\citeauthoryear{{Burgers}}{{Burgers}}{1969}]{burgers1969}
{Burgers} J.~M.,  1969, {Flow Equations for Composite Gases}.
{New York: Academic Press, 1969}

\bibitem[\protect\citeauthoryear{{Cassisi}, {Potekhin}, {Pietrinferni},
  {Catelan} \& {Salaris}}{{Cassisi} et~al.}{2007}]{cassisi2007}
{Cassisi} S.,  {Potekhin} A.~Y.,  {Pietrinferni} A.,  {Catelan} M.,
  {Salaris} M.,  2007, \apj, 661, 1094

\bibitem[\protect\citeauthoryear{{Charbonneau}}{{Charbonneau}}{1993}]{charbonn%
eau1993}
{Charbonneau} P.,  1993, \apj, 405, 720

\bibitem[\protect\citeauthoryear{{Charpinet}, {Fontaine}, {Brassard}, {Chayer},
  {Rogers}, {Iglesias} \& {Dorman}}{{Charpinet} et~al.}{1997}]{charpinet1997}
{Charpinet} S.,  {Fontaine} G.,  {Brassard} P.,  {Chayer} P.,  {Rogers} F.~J.,
  {Iglesias} C.~A.,    {Dorman} B.,  1997, \apjl, 483, L123

\bibitem[\protect\citeauthoryear{{Charpinet}, {Fontaine}, {Brassard} \&
  {Dorman}}{{Charpinet} et~al.}{1996}]{charpinet1996}
{Charpinet} S.,  {Fontaine} G.,  {Brassard} P.,    {Dorman} B.,  1996, \apjl,
  471, L103

\bibitem[\protect\citeauthoryear{{Chayer}, {Fontaine}, {Fontaine},
  {Lamontagne}, {Wesemael}, {Dupuis}, {Heber}, {Napiwotzki} \&
  {Moehler}}{{Chayer} et~al.}{2004}]{chayer2004}
{Chayer} P.,  {Fontaine} G.,  {Fontaine} M.,  {Lamontagne} R.,  {Wesemael} F.,
  {Dupuis} J.,  {Heber} U.,  {Napiwotzki} R.,    {Moehler} S.,  2004, \apss,
  291, 359

\bibitem[\protect\citeauthoryear{{Dupret}}{{Dupret}}{2001}]{dupret2001}
{Dupret} M.~A.,  2001, \aap, 366, 166

\bibitem[\protect\citeauthoryear{{Dupret}}{{Dupret}}{2003}]{dupret2003}
{Dupret} M.-A.,  2003, PhD thesis, Universit\'e de Li\`ege

\bibitem[\protect\citeauthoryear{{Edelmann}, {Heber}, {Hagen}, {Lemke},
  {Dreizler}, {Napiwotzki} \& {Engels}}{{Edelmann} et~al.}{2003}]{edelmann2003}
{Edelmann} H.,  {Heber} U.,  {Hagen} H.,  {Lemke} M.,  {Dreizler} S.,
  {Napiwotzki} R.,    {Engels} D.,  2003, \aap, 400, 939

\bibitem[\protect\citeauthoryear{{Edelmann}, {Heber} \&
  {Napiwotzki}}{{Edelmann} et~al.}{2006}]{edelmann2006}
{Edelmann} H.,  {Heber} U.,    {Napiwotzki} R.,  2006, Baltic Astronomy, 15,
  103

\bibitem[\protect\citeauthoryear{{Eggleton}}{{Eggleton}}{1971}]{eggleton1971}
{Eggleton} P.~P.,  1971, \mnras, 151, 351

\bibitem[\protect\citeauthoryear{{Eggleton}}{{Eggleton}}{1972}]{eggleton1972}
{Eggleton} P.~P.,  1972, \mnras, 156, 361

\bibitem[\protect\citeauthoryear{{Fontaine}, {Brassard}, {Charpinet}, {Green},
  {Chayer}, {Bill{\`e}res} \& {Randall}}{{Fontaine}
  et~al.}{2003}]{fontaine2003}
{Fontaine} G.,  {Brassard} P.,  {Charpinet} S.,  {Green} E.~M.,  {Chayer} P.,
  {Bill{\`e}res} M.,    {Randall} S.~K.,  2003, \apj, 597, 518

\bibitem[\protect\citeauthoryear{{Fontaine} \& {Chayer}}{{Fontaine} \&
  {Chayer}}{1997}]{fontaine1997}
{Fontaine} G.,  {Chayer} P.,  1997, in {Philip} A.~G.~D.,  {Liebert} J.,
  {Saffer} R.,   {Hayes} D.~S.,  eds, The Third Conference on Faint Blue Stars:
  {The Helium Abundance Puzzle in Subdwarf B Stars: a Theoretical
  Interpretation Through Mass Loss}.
L. Davis Press, p.~169

\bibitem[\protect\citeauthoryear{{Fontaine}, {Green}, {Chayer}, {Brassard},
  {Charpinet} \& {Randall}}{{Fontaine} et~al.}{2006}]{fontaine2006}
{Fontaine} G.,  {Green} E.~M.,  {Chayer} P.,  {Brassard} P.,  {Charpinet} S.,
   {Randall} S.~K.,  2006, Baltic Astronomy, 15, 211

\bibitem[\protect\citeauthoryear{{Formicola} \& {LUNA
  Collaboration}}{{Formicola} \& {LUNA Collaboration}}{2002}]{formicola2002}
{Formicola} A.,  {LUNA Collaboration} 2002, in {W.~Hillebrandt \&
  E.~M{\"u}ller} ed., Nuclear Astrophysics {Re-investigation of the
  $^{14}$N(p,{$\gamma$})$^{15}$O reaction within LUNA collaboration}.
p.~111

\bibitem[\protect\citeauthoryear{{Freytag} \& {Steffen}}{{Freytag} \&
  {Steffen}}{2004}]{freytag2004}
{Freytag} B.,  {Steffen} M.,  2004, in {J.~Zverko, J.~Ziznovsky, S.~J.~Adelman,
  \& W.~W.~Weiss} ed., The A-Star Puzzle Vol.~224 of IAU Symposium, {Numerical
  simulations of convection in A-stars}.
p.~139

\bibitem[\protect\citeauthoryear{{Geier}, {Heber}, {Edelmann}, {Morales-Rueda}
  \& {Napiwotzki}}{{Geier} et~al.}{2010}]{geier2010}
{Geier} S.,  {Heber} U.,  {Edelmann} H.,  {Morales-Rueda} L.,    {Napiwotzki}
  R.,  2010, \apss, 329, 127

\bibitem[\protect\citeauthoryear{{Green}, {Fontaine}, {Reed}, {Callerame},
  {Seitenzahl}, {White}, {Hyde}, {{\O}stensen}, {Cordes}, {Brassard}, {Falter},
  {Jeffery}, {Dreizler}, {Schuh}, {Giovanni}, {Edelmann}, {Rigby} \&
  {Bronowska}}{{Green} et~al.}{2003}]{green2003}
{Green} E.~M.,  {Fontaine} G.,  {Reed} M.~D.,  {Callerame} K.,  {Seitenzahl}
  I.~R.,  {White} B.~A.,  {Hyde} E.~A.,  {{\O}stensen} R.,  {Cordes} O.,
  {Brassard} P.,  {Falter} S.,  {Jeffery} E.~J.,  {Dreizler} S.,  {Schuh}
  S.~L.,  {Giovanni} M.,  {Edelmann} H.,  {Rigby} J.,    {Bronowska} A.,  2003,
  \apjl, 583, L31

\bibitem[\protect\citeauthoryear{{Grevesse} \& {Noels}}{{Grevesse} \&
  {Noels}}{1993}]{grevesse1993}
{Grevesse} N.,  {Noels} A.,  1993, in {N.~Prantzos, E.~Vangioni-Flam, \&
  M.~Casse} ed., Origin and Evolution of the Elements: {Cosmic abundances of
  the elements}.
p.~15

\bibitem[\protect\citeauthoryear{{Han}, {Podsiadlowski}, {Maxted} \&
  {Marsh}}{{Han} et~al.}{2003}]{han2003}
{Han} Z.,  {Podsiadlowski} P.,  {Maxted} P.~F.~L.,    {Marsh} T.~R.,  2003,
  \mnras, 341, 669

\bibitem[\protect\citeauthoryear{{Heber}}{{Heber}}{1986}]{heber1986}
{Heber} U.,  1986, \aap, 155, 33

\bibitem[\protect\citeauthoryear{{Heber}, {Maxted}, {Marsh}, {Knigge} \&
  {Drew}}{{Heber} et~al.}{2003}]{heber2003}
{Heber} U.,  {Maxted} P.~F.~L.,  {Marsh} T.~R.,  {Knigge} C.,    {Drew} J.~E.,
  2003, in {I.~Hubeny, D.~Mihalas, \& K.~Werner} ed., Stellar Atmosphere
  Modeling Vol.~288 of Astronomical Society of the Pacific Conference Series,
  {Stellar Wind Signatures in sdB Stars?}.
p.~251

\bibitem[\protect\citeauthoryear{{Herwig}, {Austin} \& {Lattanzio}}{{Herwig}
  et~al.}{2006}]{herwig2006}
{Herwig} F.,  {Austin} S.~M.,    {Lattanzio} J.~C.,  2006, \prc, 73, 025802

\bibitem[\protect\citeauthoryear{{Hu}, {Dupret}, {Aerts}, {Nelemans},
  {Kawaler}, {Miglio}, {Montalban} \& {Scuflaire}}{{Hu} et~al.}{2008}]{hu2008}
{Hu} H.,  {Dupret} M.,  {Aerts} C.,  {Nelemans} G.,  {Kawaler} S.~D.,  {Miglio}
  A.,  {Montalban} J.,    {Scuflaire} R.,  2008, \aap, 490, 243

\bibitem[\protect\citeauthoryear{{Hu}, {Glebbeek}, {Thoul}, {Dupret},
  {Stancliffe}, {Nelemans} \& {Aerts}}{{Hu} et~al.}{2010}]{hu2010}
{Hu} H.,  {Glebbeek} E.,  {Thoul} A.~A.,  {Dupret} M.,  {Stancliffe} R.~J.,
  {Nelemans} G.,    {Aerts} C.,  2010, \aap, 511, A87

\bibitem[\protect\citeauthoryear{{Hu}, {Nelemans}, {Aerts} \& {Dupret}}{{Hu}
  et~al.}{2009}]{hu2009}
{Hu} H.,  {Nelemans} G.,  {Aerts} C.,    {Dupret} M.,  2009, \aap, 508, 869

\bibitem[\protect\citeauthoryear{{Iglesias} \& {Rogers}}{{Iglesias} \&
  {Rogers}}{1996}]{iglesias1996}
{Iglesias} C.~A.,  {Rogers} F.~J.,  1996, \apj, 464, 943

\bibitem[\protect\citeauthoryear{{Itoh}, {Adachi}, {Nakagawa}, {Kohyama} \&
  {Munakata}}{{Itoh} et~al.}{1989}]{itoh1989}
{Itoh} N.,  {Adachi} T.,  {Nakagawa} M.,  {Kohyama} Y.,    {Munakata} H.,
  1989, \apj, 339, 354

\bibitem[\protect\citeauthoryear{{Itoh}, {Mutoh}, {Hikita} \& {Kohyama}}{{Itoh}
  et~al.}{1992}]{itoh1992}
{Itoh} N.,  {Mutoh} H.,  {Hikita} A.,    {Kohyama} Y.,  1992, \apj, 395, 622

\bibitem[\protect\citeauthoryear{{Jeffery} \& {Saio}}{{Jeffery} \&
  {Saio}}{2006}]{jeffery2006}
{Jeffery} C.~S.,  {Saio} H.,  2006, \mnras, 372, L48

\bibitem[\protect\citeauthoryear{{Kawaler} \& {Hostler}}{{Kawaler} \&
  {Hostler}}{2005}]{kawaler2005}
{Kawaler} S.~D.,  {Hostler} S.~R.,  2005, \apj, 621, 432

\bibitem[\protect\citeauthoryear{{Kilkenny}, {Koen}, {O'Donoghue} \&
  {Stobie}}{{Kilkenny} et~al.}{1997}]{kilkenny1997}
{Kilkenny} D.,  {Koen} C.,  {O'Donoghue} D.,    {Stobie} R.~S.,  1997, \mnras,
  285, 640

\bibitem[\protect\citeauthoryear{{Kupka} \& {Montgomery}}{{Kupka} \&
  {Montgomery}}{2002}]{kupka2002}
{Kupka} F.,  {Montgomery} M.~H.,  2002, \mnras, 330, L6

\bibitem[\protect\citeauthoryear{{Mendoza}, {Seaton}, {Buerger},
  {Bellor{\'{\i}}n}, {Mel{\'e}ndez}, {Gonz{\'a}lez}, {Rodr{\'{\i}}guez},
  {Delahaye}, {Palacios}, {Pradhan} \& {Zeippen}}{{Mendoza}
  et~al.}{2007}]{mendoza2007}
{Mendoza} C.,  {Seaton} M.~J.,  {Buerger} P.,  {Bellor{\'{\i}}n} A.,
  {Mel{\'e}ndez} M.,  {Gonz{\'a}lez} J.,  {Rodr{\'{\i}}guez} L.~S.,  {Delahaye}
  F.,  {Palacios} E.,  {Pradhan} A.~K.,    {Zeippen} C.~J.,  2007, \mnras, 378,
  1031

\bibitem[\protect\citeauthoryear{{Michaud} \& {Richer}}{{Michaud} \&
  {Richer}}{2008}]{michaud2008}
{Michaud} G.,  {Richer} J.,  2008, \memsai, 79, 592

\bibitem[\protect\citeauthoryear{{Michaud}, {Richer} \& {Richard}}{{Michaud}
  et~al.}{2011}]{michaud2011}
{Michaud} G.,  {Richer} J.,    {Richard} O.,  2011, \aap, 529, A60

\bibitem[\protect\citeauthoryear{{O'Toole} \& {Heber}}{{O'Toole} \&
  {Heber}}{2006}]{otoole2006}
{O'Toole} S.~J.,  {Heber} U.,  2006, \aap, 452, 579

\bibitem[\protect\citeauthoryear{{O'Toole}, {Heber}, {Chayer}, {Fontaine},
  {O'Donoghue} \& {Charpinet}}{{O'Toole} et~al.}{2004}]{otoole2004}
{O'Toole} S.~J.,  {Heber} U.,  {Chayer} P.,  {Fontaine} G.,  {O'Donoghue} D.,
   {Charpinet} S.,  2004, \apss, 291, 427

\bibitem[\protect\citeauthoryear{{Paquette}, {Pelletier}, {Fontaine} \&
  {Michaud}}{{Paquette} et~al.}{1986}]{paquette1986}
{Paquette} C.,  {Pelletier} C.,  {Fontaine} G.,    {Michaud} G.,  1986, \apjs,
  61, 177

\bibitem[\protect\citeauthoryear{{Pols}, {Tout}, {Eggleton} \& {Han}}{{Pols}
  et~al.}{1995}]{pols1995}
{Pols} O.~R.,  {Tout} C.~A.,  {Eggleton} P.~P.,    {Han} Z.,  1995, \mnras,
  274, 964

\bibitem[\protect\citeauthoryear{{Seaton}}{{Seaton}}{1993}]{seaton1993}
{Seaton} M.~J.,  1993, \mnras, 265, L25

\bibitem[\protect\citeauthoryear{{Seaton}}{{Seaton}}{2005}]{seaton2005}
{Seaton} M.~J.,  2005, \mnras, 362, L1

\bibitem[\protect\citeauthoryear{{Seaton}, {Yan}, {Mihalas} \&
  {Pradhan}}{{Seaton} et~al.}{1994}]{seaton1994}
{Seaton} M.~J.,  {Yan} Y.,  {Mihalas} D.,    {Pradhan} A.~K.,  1994, \mnras,
  266, 805

\bibitem[\protect\citeauthoryear{{Th{\'e}ado}, {Vauclair}, {Alecian} \& {Le
  Blanc}}{{Th{\'e}ado} et~al.}{2009}]{theado2009}
{Th{\'e}ado} S.,  {Vauclair} S.,  {Alecian} G.,    {Le Blanc} F.,  2009, \apj,
  704, 1262

\bibitem[\protect\citeauthoryear{{Thoul}, {Bahcall} \& {Loeb}}{{Thoul}
  et~al.}{1994}]{thoul1994}
{Thoul} A.~A.,  {Bahcall} J.~N.,    {Loeb} A.,  1994, \apj, 421, 828

\bibitem[\protect\citeauthoryear{{Turcotte}, {Richer}, {Michaud}, {Iglesias} \&
  {Rogers}}{{Turcotte} et~al.}{1998}]{turcotte1998}
{Turcotte} S.,  {Richer} J.,  {Michaud} G.,  {Iglesias} C.~A.,    {Rogers}
  F.~J.,  1998, \apj, 504, 539

\bibitem[\protect\citeauthoryear{{Unglaub}}{{Unglaub}}{2008}]{unglaub2008}
{Unglaub} K.,  2008, \aap, 486, 923

\bibitem[\protect\citeauthoryear{{Unglaub} \& {Bues}}{{Unglaub} \&
  {Bues}}{2001}]{unglaub2001}
{Unglaub} K.,  {Bues} I.,  2001, \aap, 374, 570

\bibitem[\protect\citeauthoryear{{Vink}}{{Vink}}{2004}]{vink2004}
{Vink} J.~S.,  2004, \apss, 291, 239

\bibitem[\protect\citeauthoryear{{Vink} \& {Cassisi}}{{Vink} \&
  {Cassisi}}{2002}]{vink2002}
{Vink} J.~S.,  {Cassisi} S.,  2002, \aap, 392, 553

\bibitem[\protect\citeauthoryear{{Vink}, {de Koter} \& {Lamers}}{{Vink}
  et~al.}{2000}]{vink2000}
{Vink} J.~S.,  {de Koter} A.,    {Lamers} H.~J.~G.~L.~M.,  2000, \aap, 362, 295

\bibitem[\protect\citeauthoryear{{Vink}, {de Koter} \& {Lamers}}{{Vink}
  et~al.}{2001}]{vink2001}
{Vink} J.~S.,  {de Koter} A.,    {Lamers} H.~J.~G.~L.~M.,  2001, \aap, 369, 574

\end{thebibliography}
\onecolumn
\appendix
\section[]{Solving the Burgers' equations}\label{appendix}
With the approach of \citet{thoul1994}, the Burgers' equations without radiative forces, together with the constraints of current neutrality and mass conservation, can be expressed as a matrix equation
\[
\frac{p}{K_0}\Big(
\alpha_i\frac{d\ln p}{dr}+\nu_i\frac{d\ln T}{dr}+\sum^S_{\substack{j=1\\j\neq e}}\gamma_{ij}\frac{d\ln C_j}{dr}\Big)=\sum^{2S+2}_{j=1}\Delta_{ij}W_j.
\]
Most quantities are defined as by \citet{thoul1994} except for the coefficients $\gamma_{ij}$ and $\Delta_{ij}$ which we define as
\[
\gamma_{ij}=\frac{C_i}{C}\Big(\delta_{ij}-\frac{C_j}{C}\Big)\qquad \textrm{for }i=1,...,S.
\]
Our $\gamma_{ij}$ now takes into account the He concentration gradient which was unnecessarily eliminated by \citet{thoul1994}.
The coefficients $\Delta_{ij}$ are modified to make use of  \citet{paquette1986}'s resistance coefficients ($K_{ij}$, $z_{ij}$, $z_{ij}'$ and $z_{ij}''$) derived from a screened Coulomb potential.
For $i=1,...,S$ we have
\[
\Delta_{ij}=
\begin{cases}
-\displaystyle\sum_{k\neq i} \kappa_{ik} & \textrm{for }j=i ,\\
\kappa_{ij} & \text{for }j=1,...,S\wedge j\neq i,\\
\displaystyle\sum_{k\neq i} z_{ik}\kappa_{ik}x_{ik}& \text{for } j=i+S,\\
-z_{i,j-S}\kappa_{i,j-S}y_{i,j-S}& \text{for } j=S+1,...,2S \wedge j\neq i+S.
\end{cases}
\]
For $i=S+1,...,2S$,
\[
\Delta_{ij}=
\begin{cases}
\displaystyle\sum_{k\neq j} 2.5z_{i-S,k}\kappa_{i-S,k}x_{i-S,k} & \textrm{for }j=i-S,\\
-2.5z_{i-S,j}\kappa_{i-S,j}x_{i-S,j}& \text{for }j=1,...,S \wedge j\neq i-S,\\
\displaystyle\sum_{k\neq i}[\kappa_{i-S,k}y_{i-S,k}(0.8z_{i-S,k}''x_{i-S,k}+Y_{i-S,k})-
0.4 z_{i-S,i-S}''\kappa_{i-S,i-S}] & \text{for }j=i,\\
(3+z_{i-S,j-S}'-0.8z_{i-S,j-S}'')\kappa_{i-S,j-S}y_{i-S,j-S}x_{i-S,j-S}& \text{for }j=S+1,...,2S \wedge j\neq i,
\end{cases}
\]
with $\kappa_{ij}=K_{ij}/K_0$, $x_{ij}=m_j/(m_i+m_j)$, $y_{ij}=m_i/(m_i+m_i)$ and $Y_{ij}=3y_{ij}+z_{ij}'x_{ij}m_j/m_i$. Note that the approximations of  \citet{thoul1994} are retrieved with $z_{ij}=0.6$, $z_{ij}'=1.3$ and $z_{ij}''=0.4$ as estimated from a pure Coulomb potential.

The above modifications were already made by \citet{hu2010} but not explicitly described then. In this work we adapted \citet{thoul1994}'s routine to include radiative forces. After some algebra we find that Eqs (\ref{4difeq})--(\ref{4mass}) can be written as the matrix equation
\begin{equation}\label{matrix2}
\frac{p}{K_0}\Big(-\frac{\alpha_i m_i g_{{\rm rad},i}}{k_{\rm B}T}+
\alpha_i\frac{d\ln p}{dr}+\nu_i\frac{d\ln T}{dr}+\sum^S_{\substack{j=1\\j\neq e}}\gamma_{ij}\frac{d\ln C_j}{dr}\Big) = \sum^{2S+2}_{j=1}\Delta_{ij}W_j.
\end{equation}
The terms on the left-hand side represent the contributions to the diffusion velocity by radiative levitation, gravitational settling, thermal diffusion and concentration diffusion. The matrix equation is solved by LU decomposition for each of these terms separately and the solutions are combined to give to the total diffusion velocity. In principle, we could also add the terms on the left-hand side of Eq.~(\ref{matrix2}) beforehand and solve the matrix equation in one step. However, it  can be insightful to quantify the different contributions to the diffusion velocities.

\twocolumn

\end{document}